\documentclass[graybox,natbib]{svmult}
\usepackage{newtxtext}
\usepackage{helvet}         
\usepackage{courier}        
\usepackage{type1cm}        
\usepackage{makeidx}         
\usepackage{graphicx}        
\usepackage{multicol}        
\usepackage[bottom]{footmisc}
\usepackage{amsfonts}
\usepackage{latexsym}
\usepackage{amsmath}
\usepackage{amssymb}
\usepackage{xcolor}
\usepackage{amsxtra, mathrsfs, upgreek, bm}

\usepackage{tikz}
\usetikzlibrary{calc}

\makeindex             

\DeclareMathAlphabet{\mathscr}{OT1}{pzc}{m}{it} 
\usepackage{eucal}

\newcommand{\begeq}{\begin{equation}\begin{gathered}}
\newcommand{\eqend}{\end{gathered}\end{equation}}
\newcommand{\begal}{\begin{equation}\begin{aligned}}
\newcommand{\alend}{\end{aligned}\end{equation}}

\newcommand{\R}{{\mathbb{R}}}  

\begin{document}

\title{Covering a surface with pre-stressed ribbons:  from theory to nano-structures fabrication}
\titlerunning{Covering nano-structures with pre-stressed ribbons}
\author{Alexandre Danescu, Philippe Regreny, Pierre Cremillieu, Jean-Louis Leclercq, and Ioan R. Ionescu}
\authorrunning{Danescu, Regreny, Cremillieu, Leclercq, Ionescu}
\institute{A. Danescu, Ph. Regreny, P. Cremillieu and Jean-Louis Leclercq 
\at Univ Lyon, Ecole Centrale de Lyon, CNRS, INSA Lyon, University Claude Bernard Lyon 1, CPE Lyon, CNRS, INL, UMR5270, 69130 Ecully, France \\
\email{alexandre.danescu@ec-lyon.fr, philippe.regreny@ec-lyon.fr, pierre.cremillieu@ec-lyon.fr, jean-louis.leclercq@ec-lyon.fr}
\and
Ioan R. Ionescu 
\at LSPM, University Sorbonne-Paris-Nord, 93430 Villetaneuse, France
and IMAR, Romanian Academy, Bucharest, Romania \\
\email{ioan.r.ionescu@gmail.com}
}

\maketitle

\abstract{ The paper deals with the fabrication of nano-shells from pre-stressed nano-plates release.  Due to  geometrical and technological restrictions we have to cover a given surface  with  three-dimensional thin ribbons. We  discuss the key role of the geodesic curvature  in  the design  of such shell-ribbons. We show that including small-strains but large rotations we are able to control the metric tensor of both Lagrangian and Eulerian ribbons by an appropriate choice of the width and thickness of the ribbons. Moreover, the Green-Lagrange strain tensor is controlled by the distance between the curvature of the planar ribbon and the geodesic curvature of the supporting curve of the shell-ribbon. Under suitable constitutive assumptions, we deduce the field equations, the boundary conditions and the design equations. The former relate the pre-stress in the planar layer to the final geometry of the desired shell-ribbon.  A fine tuning of the  composition, geometry and of the pre-stress of the plate-ribon is necessary to design and fabricate the shell-ribbon.
	 We design and fabricate a partial cover of the sphere with constant latitude ribbons starting from planar multi-layer semiconductor materials grown by molecular beam epitaxy.
 The details of fabrication method  and its limitations  are discussed in detail.	
}

\keywords{nonlinear elasticity, pre-stressed stuctures, shell-design, nano-structures, fabrication, epitaxial thin films}

\section{Introduction\label{intro}}

Nowadays the fabrication processes in semiconductor industry use essentially the planar technology and among the various methods of crystal growth, the molecular beam epitaxy (MBE) presents the significant advantage of highly accurate control of composition (up to $1\%$) and thickness (up to monolayer). Composition control endow multi-layered planar structures with pre-stress which may be beneficial for the design on 3D objects by pre-stress relaxation. The prototype of this phenomena is the bi-layer material where the presence of the pre-stress in one of the layers induced the bending of the free bi-layer structure. Initiated in \cite{prinz2000} (see also \cite{prinz2001}, \cite{prinz2003}, \cite{seleznev2003}, \cite{prinz2006}, \cite{prinz2017}) for simple rolls, curls and developable ribbons the method was extended to cover more complex situations in \cite{danescu2013}, \cite{danescu2018}. Introduced by an heuristic method in \cite{danescu2013} and later reconsidered in the framework of small-strains and large-rotations in \cite{danescu2021}, the geodesic curvature represents the key concept for the design of 3D structures from planar pre-stresses films. From a different point of view, the equilibirum shape of a pre-stressed material was investigated by using dimension reduction in \cite{dret_raoult_95, friesecke2002theorem, friesecke2002, friesecke2006, de2020energy, wang2019uniformly, de2019hierarchy} leading to a hierarchy of non-linear elastic models \cite{lewicka2018}.  

These previous results concerning relaxation of pre-stressed bi-layer materials focus on straight ribbons that relax toward rolls and curls, all based on isometric transformations. However, it is well-known that the class of isometries between planar and three-dimensional surfaces, extensively studied in \cite{fosdick2016}, is too narrow to cover simple non-developable surfaces occurring in pre-stressed relaxation design problems. To circumvent this theoretical drawback, in a recent paper \cite{danescu2020} we developed a shell design model built on a non-isometric perturbation assumption of Love-Kirchhoff type superposed on a plate-to-shell theory. Extending shell models in \cite{steigmann2013}, \cite{ciarlet2018}, \cite{steigmann2007thin}, \cite{steigmann2007asymptotic}, \cite{steigmann2014classical}) the geometric description involves a single small parameter $\delta\ll 1$, the product between the thickness of the shell and its curvature.  

The main difficulty in applying the shell-design model in \cite{danescu2020} is of a geometric nature. Indeed, for several common mid-surfaces the small-strain assumption drastically reduces the surface width.  However, since we are focusing on brittle-elastic materials (such as semiconductors), the small deformations assumption is merely a  technological restriction and not a mathematical simplification. To encompass this limitation, in \cite{danescu2021} another type of shell, (called a strip-shell) is constructed, for which this assumption can be fulfilled by an appropriate choice of an additional geometric parameter, namely the strip width. In this restricted framework, if the product between the strip-shell width and its curvature is of order $\delta^{1/2}$ the assumptions of plate-to-shell theory \cite{danescu2020} are fulfilled so that, for any strip of a given shell we provide a simple model able to design the corresponding plate-strip (i.e., to compute the shape and pre-stress momentum of the plate). The next step analysed here is to cover the given surface (shell) with one or several strips, situation in which we can provide an explicit design of the corresponding planar (plate) strips. 

The paper is organized as follows : the first two sections recall the geometric and mechanical  assumptions of the plate to shell model for design proposed in \cite{danescu2020}. We relate the geometric aspects to the pre-stress via constitutive relations and field equations in finite strains through the assumption of weak-transversal heterogeneity, assumption fulfilled here by the weak variation of the composition in our crystal growth process. The third section discuss the main geometric aspects of the theory (see \cite{danescu2021} for more details), with a particular accent on the metric tensors for planar ribbons along curves and three dimensional ribbons as subsets of arbitrary surfaces in ${\mathbb R}^3.$ The main result shows that the distance between the curvature of the planar curve (the planar design) and the geodesic curvature of the three dimensional supporting curve of the ribbon controls the Green-Lagrange strain tensor, so that the small-strain (but large rotations) assumptions can be fulfilled by an appropriate choice of the planar geometry. The fourth section descibes a specific application: fabrication of a partial cover of the sphere from a planar pre-stressed bilayer material by using a design based on the geodesic curvature of constant-lattitude circles.    

\section{Geometric and kinematical settings }

Let us consider a plate with mid-surface $R_0 \subset \R^2$ and thickness $H=H(\bar{X})$ in the Lagrangian configuration (here $\bar{X}=(X_1,X_2)$) and let $S_0 \subset \R^3$ be the  mid-surface  of an  Eulerian  shell  of thickness $h$,  with  $\bm{e}_3$ the  unit normal  and  ${\cal K}$ the curvature tensor acting from the tangent plane into itself. 

In what follows, $\delta \ll 1$ will be a small parameter characterizing the Eulerian and Lagrangian shell thickness and such that : 
\begin{equation}
h\vert{\cal K}\vert = {\cal O}(\delta),\qquad H/L_c = {\cal O}(\delta), \quad \vert \nabla_2 H\vert ={\cal O}(\delta), 
\end{equation}
where $L_c$ is the characteristic length of the surface and $\nabla_2$ is the gradient with respect to $\bar{X}\in  R_0$. 

The  main geometric assumption in \cite{danescu2020}  is that there exists a transformation $\bm{x} : R_0 \to S_0$   of  the Lagrangian mid-surface $R_0$ into the designed  Eulerian one $S_0$  such that the associated deformation of the geometric transformation is small, i.e., 
\begin{equation} \label{AG}
\vert \bm{E}_2 \vert=\frac{1}{2}\vert \nabla_2^T \bm{x}\nabla_2 \bm{x}-\bm{I}_2\vert = \mathcal{O}(\delta).
\end{equation}
Here   $\bm{I}_2=\bm{c}_1\otimes\bm{c}_1+\bm{c}_2 \otimes\bm{c}_2$ is the identity tensor on $\R^2$ and $\{ \bm{c}_1,\bm{c}_2, \bm{c}_3\}$ is the Cartesian basis in the Lagrangian description and herafter we denote by $\mathbf{K}=\nabla_2^T \bm{x}{\mathcal{K}}\nabla_2 \bm{x}$  the Lagrangian curvature tensor acting from $\R^2$ into itself. 

The kinematics of the plate deformation considered in \cite{danescu2020}  involves the classical Love-Kirchhoff assumption, i.e.: {\em the normal to the plate mid-plane remains normal to the designed mid-surface} but in a finite deformation context and thus including large rotations. In addition, the transversal deformation is affine with respect to the plate thickness. Superposed to  the kinematics associated to the exact design which reproduces the target mid-surface, we  consider a small perturbation of Love-Kirchhoff type in order to compensate the small (membrane) deformation of the proposed geometric transformation. As a consequence, the mid-surface of this overall kinematics will be close to the designed mid-surface, and for this reason we called it {\em approximate design kinematics}. 

\section{Weak transversal homogenenity and the moment equations}

Although, the general theory developed in \cite{danescu2020} can cover the general anisotropic framework, here we restrict to cubic materials since our designed experiment involve multilayered cubic III-V semiconductor alloy $\textrm{In}_{1-\alpha}\textrm{Ga}_\alpha\textrm{P}$ for $\alpha$ small. In order to account for small-strains but large rotations including inhomogeneous pre-stress we consider a linear constitutive relation between the second Piola-Kirchhoff stress $\mathbf{S}$ tensor and the Green strain-tensor $\mathbf{E}=\frac{1}{2}(\mathbf{F}^T\mathbf{F}-\mathbf{I})$ in the form

\begin{equation}
	\mathbf{S}={\mathbb C}(X_3)[\mathbf{E}]+\mathbf{S}^\star(X_3)+\Sigma{\cal O}(\delta^2),
\end{equation}
where $\Sigma$ is a characteristic stress and both the material parameters $\mathbb{C}=(C_{ij})$ (Voigt notation) and the pre-stress $\mathbf{S}^\star$ depends on the normal coordinate in the reference configuration. Morover, following \cite{danescu2021} we assume that the elasticities ${\mathbb C}$ obey the {\em weak transversal heterogeneity} condition, i.e.,
\begin{equation}
	\langle{C}_{ij}\rangle_2 =\Sigma{\cal O}(\delta),\qquad \langle{C}_{ij}\rangle_3=\frac{1}{12}\langle{C}_{ij}\rangle_1+\Sigma{\cal O}(\delta),
\end{equation}
where the successive averages (moments) of a $X_3$-dependent function $\langle{f}\rangle_n$ ($n=1,2,3)$ are defined through
\begin{equation}
	\langle{f}\rangle_n=\frac{1}{H^{n}}\int_{-H/2}^{H/2}X_3^{n-1}{f}(X_3) dX_3.
\end{equation}

Taking into account that during the MBE growth the upper surface of the multi-layer structure is  stress-free, we assume that the pre-stress acting surfaces parallel to the mid-surface vanishes, so that $\mathbf{S}^\star\mathbf{e}_3=\mathbf{0}.$ Then, following \cite{danescu2020}, the moments equations are
\begin{equation}\label{field}
	\textrm{div}(\frac{H}{12}\mathbb{M}[\mathbf{K}]+\langle\mathbf{S}^\star_2\rangle_2)=\mathbf{0}\ \textrm{in}\ R_0,\quad
	(\frac{H}{12}\mathbb{M}[\mathbf{K}]+\langle\mathbf{S}^\star_2\rangle_2)\mathbf{\nu}_{ext}=\mathbf{0}\ \textrm{on}\ \partial R_0,
\end{equation}
\begin{equation}\label{condition}
	(\frac{H}{12}\mathbb{M}[\mathbf{K}]+\langle\mathbf{S}^\star_2\rangle_2):\mathbf{K}=\mathbf{0}\quad \textrm{in}\ R_0,
\end{equation}
where $\mathbf{S}^\star_2$ is the in-plane pre-stress  and $\mathbb{M}=\{M_{ij}\}$ is related to the in-plane reduced elasticity, i.e.,
\begin{equation}
 \mathbb{M}[\bm{A}]=\langle \mathbb{D}_2\rangle_1[\bm{A}]-\frac{\langle C_{12}\rangle_1^2}{\langle C_{11}\rangle_1}(\bm{I}:\bm{A})\bm{I}
\end{equation}
and  $\mathbb{D}_2$ is the in-plane part of the Voigt tensor.

Obviously, equations (\ref{field})-(\ref{condition}) are satisfied if the pre-stress $\bm{S}^\star$ is such that 
\begin{equation}\label{prestress_curvature}
	\langle\bm{S}^\star_2\rangle_2=-\frac{H}{12}\mathbb{M}[\bm{K}].
\end{equation}

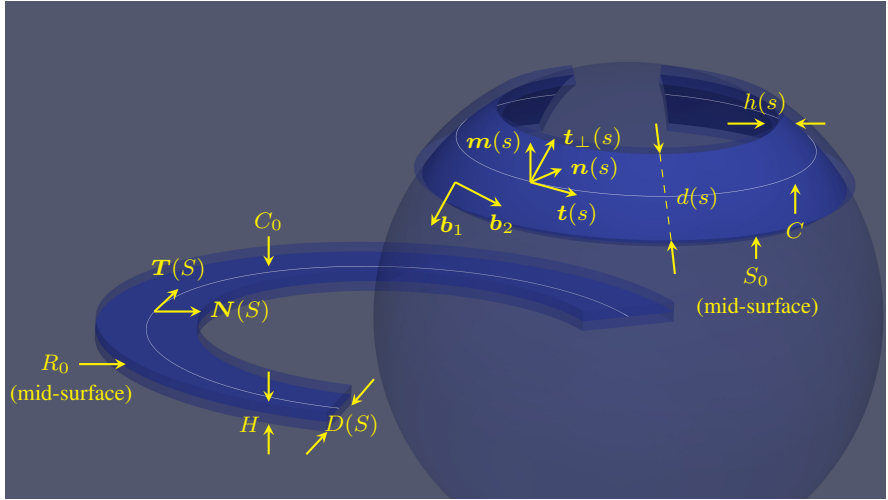
\begin{figure}[ht!]
\centering
\begin{tikzpicture}[scale=1]
    \node[anchor=south west,inner sep=0] at (0,0) 
    {\includegraphics[width=\textwidth]{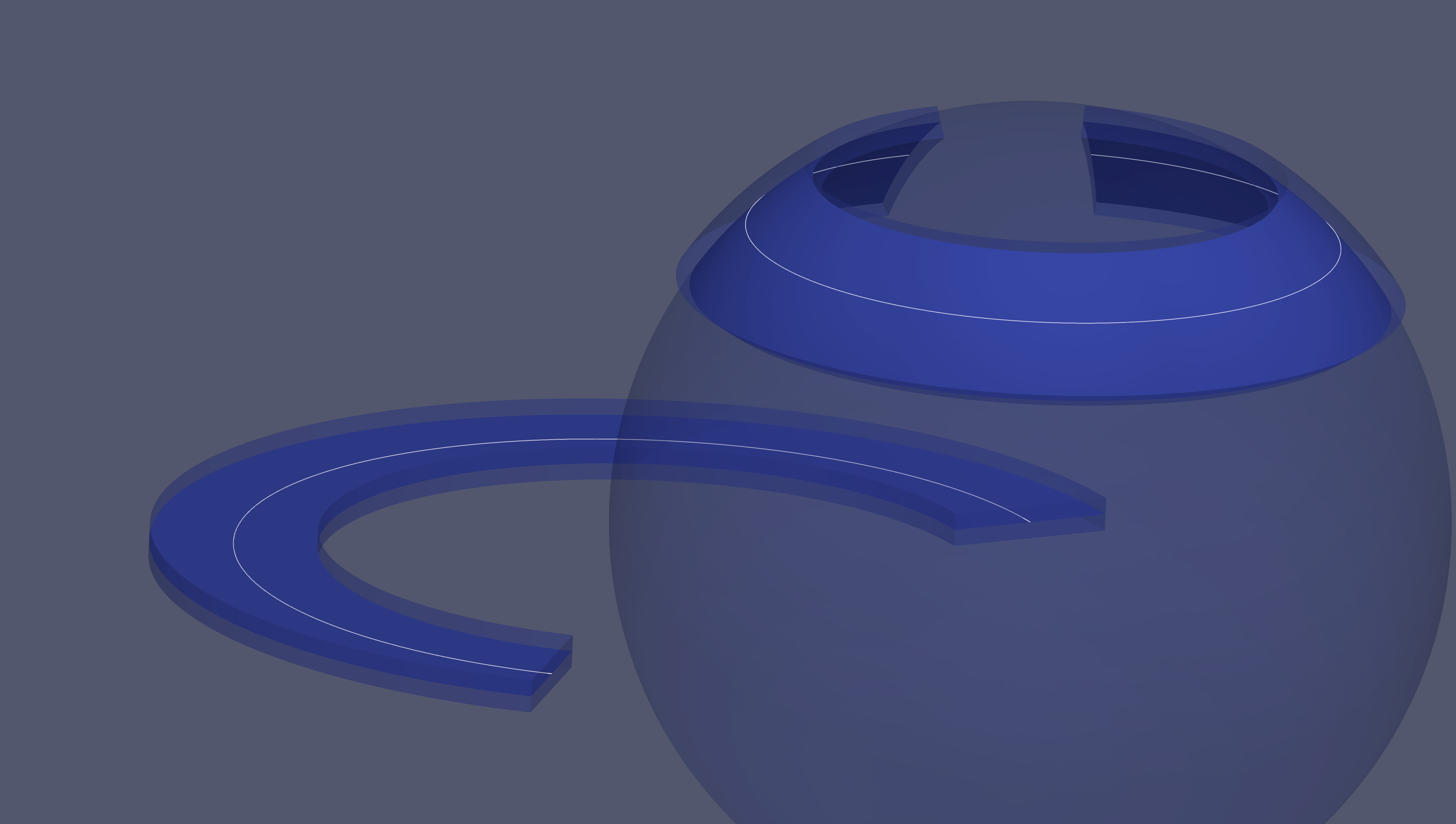}};
    \draw[->,>=stealth,color=yellow,thick] (3.5,1.7) -- (3.5,1.3);
    \draw[->,>=stealth,color=yellow,thick] (3.5,0.6) -- (3.5,1.) node[left] {$H$};
    \draw[->,>=stealth,color=yellow,thick] (4.9,1.6) -- (4.6,1.25) node[left,below] {$D(S)$};
    \draw[->,>=stealth,color=yellow,thick] (4.0,0.6) -- (4.28,0.9);
    \draw[->,>=stealth,color=yellow,thick] (1.98,2.5) -- (2.3,2.8) node[left,above] {$\bm{T}(S)$};
    \draw[->,>=stealth,color=yellow,thick] (1.98,2.5) -- (2.6,2.5) node[right] {$\bm{N}(S)$};
    \draw[<-,>=stealth,color=yellow,thick] (1.6,1.8)  -- (0.98,1.8) node[left] {$R_0$};
    \draw[color=yellow,thick] (1.8,1.4) node[left] {(mid-surface)};
    \draw[->,>=stealth,color=yellow,thick] (3.5,3.5) -- (3.5,3.1);
    \node[color=yellow] at (3.5,3.7) {$C_0$};
    
    \draw[->,>=stealth,color=yellow,thick] (6.98,4.22) -- (7.3,4.8) node[right] {$\bm{t}_{\perp}(s)$};
    \draw[->,>=stealth,color=yellow,thick] (6.98,4.22) -- (7.6,4.05) node[below] {$\bm{t}(s)$};
    \draw[->,>=stealth,color=yellow,thick] (6.98,4.22) -- (7.4,4.4) node[right] {$\bm{n}(s)$};
    \draw[->,>=stealth,color=yellow,thick] (6.98,4.22) -- (6.98,4.75) node[left] {$\bm{m}(s)$};
    \draw[<-,>=stealth,color=yellow,thick] (9.98,3.5) -- (9.98,3.2) node[below] {$S_0$};
    \draw[<-,>=stealth,color=yellow,thick] (10.5,4.2) -- (10.5,3.8) node[below] {$C$};
    \draw[color=yellow,thick] (10,2.8) node[below] {(mid-surface)};
    \draw[<-,>=stealth,color=yellow,thick] (10.5,5.) -- (10.9,5.);
    \draw[->,>=stealth,color=yellow,thick] (9.6,5.) -- (10.1,5.) node[above] {$h(s)$};
    \draw[->,>=stealth,color=yellow,thick] (8.9,3) -- (8.85,3.45);
    \draw[->,>=stealth,color=yellow,thick] (8.65,5) -- (8.7,4.6);
    \draw[-,>=stealth,color=yellow,thin,dashed] (8.7,4.6) -- (8.85,3.45);
    \node[color=yellow] at (9.2,4.0) {$d(s)$};
    \draw[->,>=stealth,color=yellow,thick] (5.98,4.22) -- (5.66,3.64) node[right] {$\bm{b}_{1}$};
    \draw[->,>=stealth,color=yellow,thick] (5.98,4.22) -- (6.6,3.9) node[below] {$\bm{b}_2$};
\end{tikzpicture}
	\caption{Geometric elements of the planar ribbon : thickness $H(S),$ width $D(S),$ the tangent and normal vectors ($\bm{T}(S)$ and $\bm{N}(S)$) along the curve $C_0$ located in the mid-surface $R_0$ and the geometric elements for the shell-ribbon with mid-surface $S_0$ along the curve $C$: $\{ \bm{t}(s),\bm{n}(s),\bm{m}(s)\}$ Frenet frame along $C,$ the vector $\bm{t}_\perp$ (located in the tangent plane to $S_0$), thickness $h(s)$ and width $d(s).$}
	\label{geometric_elements}
\end{figure}

 \section{Small strain deformation of a ribbon}

If $S\in[0,L]$ and $\kappa_0(S)$ are the arc-length and the curvature of a planar curve $C_0$ located at $\bm{R}(S)\in \mathbb{R}^2$ with tangent $\bm{T}(S)$ and normal $\bm{N}(S),$ we define (see Figure \ref{geometric_elements})  the planar ribbon  $R_0\subset \R^2$ along $C_0$ of  width $D=D(S)$ as 
\begin{equation}
R_0=\{\bm{R}(S)+Q\bm{N}(S); S\in(0,L),Q\in(-D(S),D(S))\}.
\label{planar_ribbon}
\end{equation}

Let $S_0$  denotes the mid-surface of a shell,   given by its parametric description $\bm{u} \to \bm{r}_{S_0}(\bm{u}) \in \R^3$, where $\bm{u}=(u_1,u_2)$ are coordinates in some subset $\Omega \subset \R^2$. If  $\delta\ll 1$ is a small parameter, our goal is to provide conditions for the existence of a map $\bm{x}:R_0\rightarrow S_0$  with small strain, i.e. (\ref{AG}) holds 
uniformly with respect to $\bm{X}\in R_0$.  

Let us compute the (Lagrangian) metric tensor of the planar ribbon defined in (\ref{planar_ribbon}).  The local basis, associated to the coordinates $(S,Q)$,  is $\bm{b}_S=\bm{T}-Q\kappa_0\bm{N}$ and $\bm{b}_Q=\bm{N}$ and thus the Lam\'e coefficients and metric tensor are
\begin{equation}\label{metric_reference}
  L_S^2=g_{SS}=1-2Q\kappa_0+Q^2\kappa_0^2,\qquad g_{SQ}=0,\qquad L_Q^2=g_{QQ}=1.
\end{equation}

In order to chose among the multiple ways that map a ribbon on a surface, we study the particular case in which the ribbon cover the shell mid-surface along a given curve $C\subset S_0$ (see Figure \ref{geometric_elements}).
As a curve in $\mathbb{R}^3,$ $C$ posses its intrinsic geometric features : arc-length $s,$ Frenet frame $(\bm{t}(s),\bm{n}(s),\bm{m}(s)),$ curvature $\kappa(s)$ and torsion $\tau(s)$ and, obviously, the tangent plane to the shell  mid-surface $S_0$ contains the tangent vector $\bm{t}(s)$ to $C.$
 
Let $\bm{u}^0(s)=(u_1^0(s),u_2^0(s))$ be the arc-length parametrization of the curve $C\subset S_0.$ Then, $\bm{t}=\frac{\partial \bm{r}}{\partial u_i^0}\frac{du_i^0}{ds}=\bm{b}_i \frac{du_i^0}{ds}$ is the description of the tangent vector to $C$ in the covariant basis $\{\bm{b}_1,\bm{b}_2\}$ on $S_0.$
The main idea is to map the $q$ coordinates in a neighborhood of the curve $C\subset S_0$ in the direction $\bm{t}_\perp(s),$ which is orthogonal to its tangent vector  of the curve and belongs to the tangent plane of the surface , i.e.   $\bm{t}_\perp(s)=\bm{e}_3(\bm{u}^0(s)) \wedge \bm{t}(s)$. More precisely, if we put  
\begin{equation}
  u_i(s,q)=u_i^0(s)+q v_i(s,q),  \quad  v_i^0(s)=v_i(s,0), \quad i=1,2,
 \end{equation}
 then  the ribon surface is given by 
\begin{equation} \label{ribbon}
S_0=\{ \bm{r}_{S_0}(\bm{u}(s,q)) ; s\in(0,l),q\in(-d(s),d(s))\},
\end{equation}
 where $d$ is the ribbon width,  and  from 
$\bm{t}_\perp(s)=\frac{\partial\bm{r}}{\partial q}(s,0)=\frac{\partial \bm{r}}{\partial u_i}\frac{\partial u_i}{\partial q}(s,0)=\bm{b}_i v_i^0(s)$ we get
 \begin{equation}\label{serie}
 v_i^0(s)=\bm{t}_\perp(s)\cdot \bm{b}^i (\bm{u}^0(s)).
 \end{equation}

 If the order of magnitude for the ribbons widths with respect to the curvatures of the curves $C_0$ and $C$ as well as to the curvature tensor of the surface $S_0,$ are such that :
 \begin{equation}\label{metric_conditions}
 	D(S)\kappa_0(S)={\cal O}(\delta^{1/2}),\quad   d(s)\vert{\cal K}(\bm{u}^0(s))\vert,\quad   
 	d(s)\kappa(s)(\bm{t}_{\perp}\cdot\bm{m})={\cal O}(\delta^{1/2}),
 \end{equation}
then  an estimation of  the Lagrangian and Eulerian metric tensors  at order ${\cal O}(\delta)$ was  obtained in \cite{danescu2021}. To see that, let us compute the (Eulerian) metric tensor of the surface $S_0$ up to first-order with respect to $q.$ We have successively
 \begin{equation} \nonumber
 \begin{split}
 & g_{qq}(s,q) = g_{ij}(s,q)\frac{\partial u_i}{\partial q}\frac{\partial u_j}{\partial q}=\left(g_{ij}(s,0)+q\frac{\partial g_{ij}}{\partial q}(s,0)\right)\frac{\partial u_i}{\partial q}\frac{\partial u_j}{\partial q}+  \mathcal{O}(\delta)= \\ &g_{ij}(s,0) v_i^0 v_j^0+ q \left( 
  4g_{ij}(s,0) v_i^0\frac{\partial v_j}{\partial q}(s,0) + 
  \frac{\partial g_{ij}}{\partial u_k}(s,0) v_i^0 v_j^0 v_k^0\right) + \mathcal{O}(\delta)\\ &=\vert \bm{t}_\perp(s)\vert^2 + qv_i^0(s)
  \left( 
  4g_{ij}(s,0)\frac{\partial v_j}{\partial q}(s,0)+
  \frac{\partial g_{ij}}{\partial u_k}(s,0)v_j^0(s) v_k^0(s)\right)+ \mathcal{O}(\delta), 
 \end{split}
 \end{equation}
 and by choosing 
 \begin{equation}\label{dotv}
 \frac{\partial v_l}{\partial q}(s,0)=-\frac{1}{4}g^{li}(s,0)\frac{\partial g_{ij}}{\partial u_k}(s,0) v_j^0(s)v_k^0(s),
 \end{equation}
 we obtain $g_{qq}=1+ \mathcal{O}(\delta).$ Moreover, 
 \begin{equation} \nonumber
 \begin{split}
  g_{sq} & = g_{ij}\frac{\partial u_i}{\partial s}\frac{\partial u_j}{\partial q}=
  \left(g_{ij}(s,0)+q\frac{\partial g_{ij}}{\partial q}(s,0)\right)\frac{\partial u_i}{\partial s}
  \frac{\partial u_j}{\partial q}+ \mathcal{O}(\delta)=\\ & =g_{ij}(s,0)\frac{du_i^0}{ds}v_j^0 + 
q \left(
  g_{ij}(s,0)\frac{d v_i^0}{d s}v_j^0 + 2g_{ij}(s,0)\frac{du_i^0}{ds}\frac{\partial v_j}{\partial q}(s,0)+ \right. \\ & \left. + \frac{\partial g_{ij}}{\partial u_k}(s,0) v_k^0\frac{du_i^0}{ds}v_j^0\right)+ \mathcal{O}(\delta)
 \end{split}
 \end{equation}
 and, since $g_{ij}(s,0)\frac{du_i^0}{ds}v_j^0=\bm{t}\cdot\bm{t}_\perp=0,$ using  (\ref{dotv}) we obtain
 \begin{equation}  \nonumber
 \begin{split}
 g_{sq}& = \frac{q}{2} \left(
  2 g_{ij}(s,0)\frac{d v_i^0}{d s}v_j^0 + \frac{\partial g_{ij}}{\partial u_k}(s,0)\frac{du_i^0}{ds}v_k^0 v_j^0
  \right) + \mathcal{O}(\delta) = \\ &= \frac{q}{2}\frac{d}{ds}\left(\bm{t}_\perp\cdot\bm{t}_\perp\right)+ \mathcal{O}(\delta).
  \end{split}
 \end{equation}
 
Finally, 
\begin{equation} \nonumber
\begin{split}
 g_{ss}(s,q) & =\left(g_{ij}(s,0)+q\frac{\partial g_{ij}}{\partial u_k}\frac{\partial u_k}{\partial q}\right)\frac{\partial u_i}{\partial s}\frac{\partial u_j}{\partial s} + \mathcal{O}(\delta)=
 g_{ij}(s,0)\frac{du_i^0}{ds}\frac{du_j^0}{ds}+\\
 & +q \left(
 2 g_{ij}(s,0)\frac{du_i^0}{ds}\frac{d v_j^0}{d s} +\frac{\partial g_{ij}}{\partial  u_k}(s,0)v_k^0\frac{du_i^0}{d s}\frac{du_j^0}{ds}\right) + \mathcal{O}(\delta)= \\ &= \vert \bm{t}(s)\vert^2 +2q \left(\frac{d\bm{t}_\perp}{ds}\cdot\bm{t}\right)+ \mathcal{O}(\delta)=1+2 q \left(\frac{d\bm{t}_\perp}{ds}\cdot\bm{t}\right) + \mathcal{O}(\delta).
\end{split}
\end{equation}
But, since $\bm{t}_\perp\cdot\bm{t}=0$ we have $\bm{t}_\perp = (\bm{t}_\perp\cdot\bm{n})\bm{n}+(\bm{t}_\perp\cdot\bm{m})\bm{m}$ so that, using the Frenet formulae, we obtain
\begin{equation}
 \frac{d\bm{t}_\perp}{ds}\cdot\bm{t}=-\frac{d\bm{t}}{ds}\cdot\bm{t}_\perp=-\kappa (\bm{t}_\perp\cdot\bm{m}).
\end{equation}
This last  result emphasize the role played by the {\em geodesic curvature}  $\kappa^{geo}=\kappa (\bm{t}_\perp\cdot\bm{m})$, which is the projection of the curvature of $C$ into the tangent plane of the manifold $S_0$, in the estimation of the metric tensor.  To summarize, we obtained :
\begin{equation}\label{metric_deformed}
 g_{ss}=1-2q\kappa(s)\bm{t}_\perp\cdot\bm{m}+ \mathcal{O}(\delta),\quad 
 g_{sq}=\mathcal{O}(\delta),\quad 
 g_{qq}=1+ \mathcal{O}(\delta),
\end{equation}
\begin{equation}\label{metric}
	g_{SS}=1-2Q\kappa^0(S)+ \mathcal{O}(\delta),\quad 
	g_{SQ}=0,\quad 
	g_{QQ}=1+ \mathcal{O}(\delta).
\end{equation}

By using $(s,q)=(S,Q)\in(0,L)\times(-D,D),$ we are now able to  estimate the Green-Lagrange strain tensor  of the map $\bm{x}:R_0\rightarrow S_0$.  Since the gradient tensor  $\bm{F}$ can be written as $\bm{F}=\bm{b}_s\otimes\bm{b}^S+\bm{b}_q\otimes\bm{b}^Q,$ taking into account (\ref{metric}) and (\ref{metric_deformed}), we obtain 
$$\bm{F}^T\bm{F}=L_s^2\bm{b}_S\otimes\bm{b}_S+\bm{b}_Q\otimes\bm{b}_Q+ \mathcal{O}(\delta)=\bm{I}-(1-g_{ss}/g_{SS})\bm{e}_S\otimes\bm{e}_S+ \mathcal{O}(\delta)$$ 
and thus
\begin{equation}
\begin{split}
\bm{E}_2 & =\frac{1}{2}(\frac{g_{ss}}{g_{SS}}-1)\bm{e}_S\otimes\bm{e}_S+ \mathcal{O}(\delta)=\\ 
         & =Q[\kappa_0(S)-\kappa(S)(\bm{t}_\perp\cdot\bm{m})]\bm{e}_S\otimes\bm{e}_S+ \mathcal{O}(\delta).
         \end{split}
\label{design}
\end{equation}
We conclude that by choosing the curvature of the planar curve $C_0$ equal to the geodesic curvature of the supporting curve of the shell-ribbon $C\subset S_0$ the Green-Lagrange tensor is small, i.e.
\begin{equation} \label{KK}
	\mbox{if}\quad  \kappa_0=\kappa^{geo}=\kappa\bm{t}_\perp\cdot\bm{m},\qquad \mbox{then} \quad \bm{E}=\mathcal{O}(\delta).
\end{equation}

\section{  From theory to fabrication  of a  nano-sphere}
\label{subsection_sphere}

Let $(r, \theta, \phi)$ be the spherical coordinates in ${\mathbb R}^3$ and denote by  $\bm{e}_r=\bm{e}_r(\theta,\phi),$ $\bm{e}_\theta=\bm{e}_\theta(\theta,\phi),$ $\bm{e}_\phi= \bm{e}_\phi(\phi)$ the local physical basis. Let $S_0$ denote the surface of the sphere of radius $R_*$ with Lam\'e coefficients $L_\theta=R_*, L_\phi=R_*\sin(\theta)$ and the unit normal   $\bm{e}_3(\theta, \phi)=\bm{e}_r(\theta,\phi)$. Then, the curvature tensor on $S_0$ is $${\cal K}=-\frac{1}{R_*}\left(\bm{e}_\theta\otimes\bm{e}_\theta+ \bm{e}_\phi\otimes\bm{e}_\phi\right).$$
Let $C\subset S_0$ be a given curve with arc-length $s,$ parametric description $s\to (\theta^0(s),\phi^0(s))$ and geodesic curvature $\kappa^{geo}(s).$ If $C_0$ is the planar curve with curvature $\kappa_0(s)=\kappa^{geo}(s)$ and $R_0$ is the planar ribbon along $C_0$ (see definition (\ref{planar_ribbon})) with the width $d(s)$ such that (\ref{metric_conditions}) holds then, from the small-strain membrane condition (\ref{AG}) we get 
$$\displaystyle \bm{K}=\frac{1}{R_*}( \bm{I}_2+{\cal O}(\delta)).$$  

From the  plate-to shell model  we find that a shell-ribbon $S_0$ of a spherical shell of radius $R_*$ along the curve $C$, could  be designed from a planar ribbon $R_0$  along a curve $C_0$  if  (\ref{KK}) holds. The pre-stress momentum have to be designed such that 
	$\displaystyle  \langle\bm{S}^*_2\rangle_2=-\frac{H^3}{12R_*}\mathbb{M}[\bm{I_2}]$  
	and can be obtained with an isotropic and homogeneous pre-stress, i.e., $\bm{S}^*_2=\sigma^* \bm{I}_2$, where 
\begin{equation}\label{RRs}
	\langle\sigma^*\rangle_2=\frac{H}{12R_*}
	\frac{\langle C_{11}^2\rangle_1+\langle C_{12}\rangle_1\langle C_{11}\rangle_1-2\langle C_{12}\rangle_1^2}{\langle C_{11}\rangle_1}.
\end{equation}

\subsection{Optimal covering with constant parallel ribbons}

For constant latitude curves, i.e.,  $\theta(s)=\theta^0,$ we have $\phi^0=s/(R_*\sin(\theta_0))$ so that $d/R_*={\cal O}(\delta^{1/2}),$  $d\cot(\theta^0)/R_*={\cal O}(\delta^{1/2})$ and the goedesic curvature is $\kappa_0=\kappa^{geo}=\cot(\theta^0)/R_*.$ This means that the width of successive ribbons will decrease with the latitude. As a straightforward consequence, the fit of successive positions and widths of constant latitude ribbons for a complete cover of the sphere is a nontrivial problem. We recall here a result from \cite{danescu2021} concerning a semi-analytical optimal covering of the sphere. 

\begin{figure}[ht!]
\centering
\includegraphics[width=0.9 \textwidth]{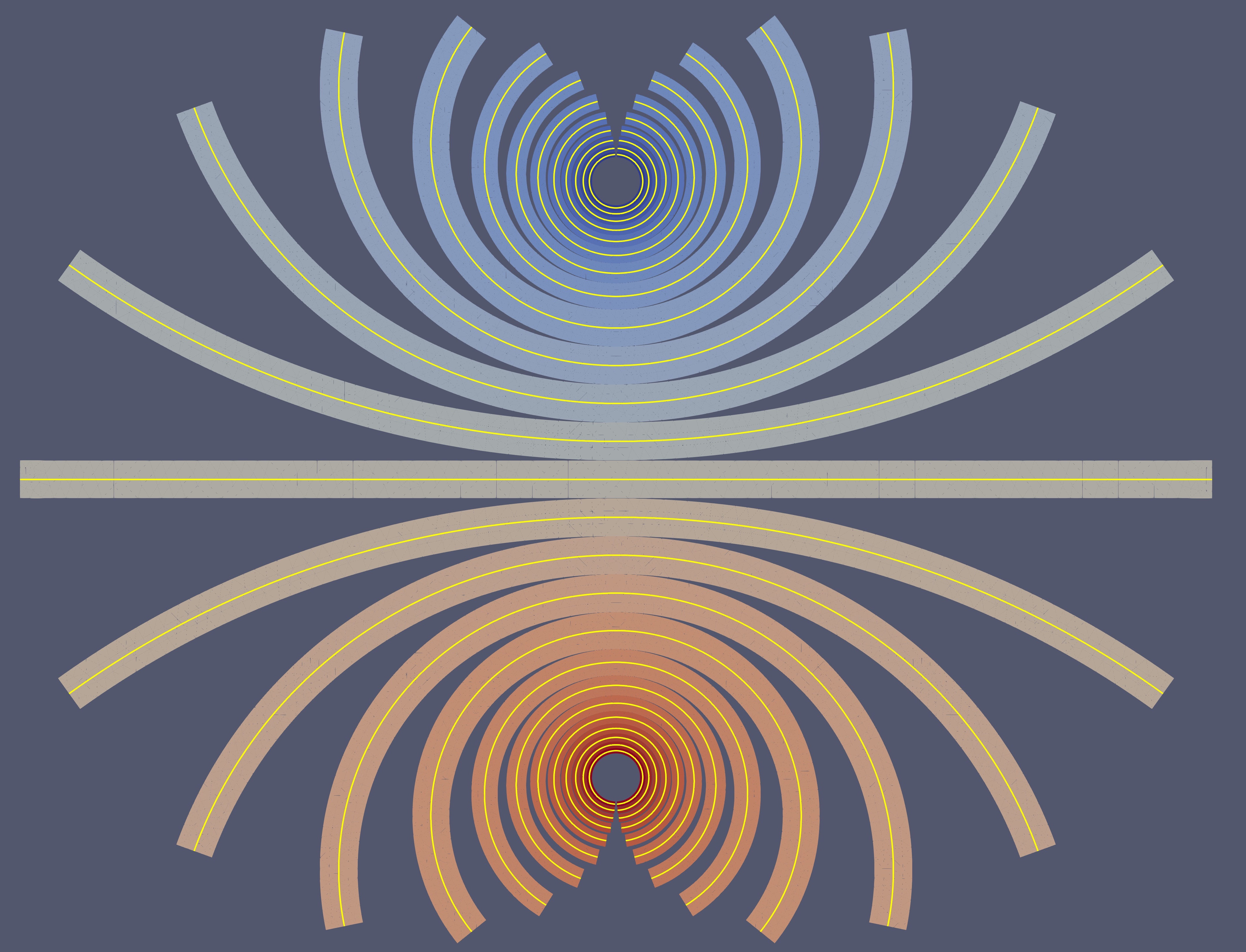}
\caption{An optimal  covering of the sphere with constant latitude ribbons obtained by implementing the solutions of  recursive system (\ref{recursive}) for $\delta=10^{-2}$.}
\label{design_full}
\end{figure}

If $\theta_k$ denote the latitudes of the supporting curve for successive ribbons then, for the $k^{\textrm{th}}$ ribbon, the arc-length is such that $s\in(-\pi R_* \cos\theta_k, \pi R_* \cos\theta_k)$ and the angular variable $\theta(q)=\pi/2-\theta_k-q/R_*$ for $q\in(-d_k,d_k).$ A symmetric solution can be obtained as follows : take $\theta_0=0,$  $\theta_{_k}=-\theta_k$ and notice that the covering condition and (\ref{metric_conditions}) can be expressed as 
\begin{equation}
 d_k\leq \delta^{1/2} R_* \min(1,\cot\theta_k),\qquad R_*(\theta_{k+1}-\theta_k)=d_k+d_{k+1}.
\end{equation}
It follows that for $\theta<\pi/4$ we can consider constant width ribbons with $\theta_k=2k\delta^{1/2}$ and thus $d_k=\delta^{1/2} R_*$ for $\vert k\vert\leq \lfloor{\frac{\pi}{8\sqrt{\delta}}-\frac{1}{2}}\rfloor$ (here $[x]$ is the entire part of $x$) while for $k> \lfloor{\frac{\pi}{8\sqrt{\delta}}-\frac{1}{2}}\rfloor$  we have to solve recursively the nonlinear equation
\begin{equation}\label{recursive}
 x-\delta^{1/2}\cot x = \theta_k+\frac{d_k}{R_*},
\end{equation}
whose solution $\theta_{k+1}$ provide 
\begin{equation}
d_{k+1}=R_*(\theta_{k+1}-\theta_k).
\end{equation} 
An implementation of this procedure for $\delta=10^{-2}$ provide the design illustrated in Figure \ref{design_full}.

\subsection{Elastic layers with pre-stress : material parameters}

 The experimental implementation of the sphere covering with variable width ribbons, presented in the previous subsection, is difficult due to very sharp angles between successive ribbons near the vertical symmetry axis, and thus incompatible with the spatial resolution of the photo-lithography processes. However, in order to illustrate the role of the geodesic curvature in the design problem we focus here on the partial cover of the sphere with constant latitude ribbons.

Since the planar design is dependent on the target surface curvature we start by the epitaxial growth of the bi-layer semiconductor structure : a 60 nm thick In$_{0.88}$Ga$_{0.12}$P layer (further denoted InGaP) grown on a 145 nm thick InP layer. The bi-layer was grown on an InP substrate previously covered by a 500 nm thick InGaAs sacrifical layer. Using data from the litterature, we have :
\begin{equation}
\begin{array}{lcrclcr}
C^{\textrm{InP}}_{11}&=& 101.1\ \textrm{GPa},&\qquad &C^{\textrm{InGaP}}_{11}&=&105.8\ \textrm{GPa},\cr
C^{\textrm{InP}}_{12}&=&  56.1\ \textrm{GPa},&\qquad &C^{\textrm{InGaP}}_{12}&=& 56.8\ \textrm{GPa},\cr C^{\textrm{InP}}_{44}&=&  45.6\ \textrm{GPa},&\qquad &C^{\textrm{InGaP}}_{44}&=& 48.5\ \textrm{GPa},
\end{array} 
\end{equation}
so that the caracteristic stress $\Sigma=100$ GPa. For $\delta=10^{-2}$ we verify that indeed 
\begin{equation}
 \langle C_{ij}\rangle_2 = \Sigma{\cal O}(\delta),\qquad \langle {C}_{ij}\rangle_3-\frac{1}{12}\langle {C}_{ij}\rangle_1=\Sigma{\cal O}(\delta),
\end{equation}
so that the assumption of weak transversal homogeneity is fullfilled. The lattice parameters for the InP and InGaP layers are respectively
\begin{equation}
 a_{\textrm{InP}}=5.8687\ \text{\normalfont{\AA}},\qquad
 a_{\textrm{In}_{0.88}\textrm{Ga}_{0.12}\textrm{P}}=5.8185\ \text{\normalfont{\AA}},
\end{equation}
and correspond to a spherical pre-strain (extension) in the upper layer of $\bm{m}=\textrm{diag}(0.86\%, 0.868\%).$ For pratical applications, it is the fraction of $\textrm{Ga}$ in the upper layer ($\textrm{In}_\alpha\textrm{Ga}_{1-\alpha}\textrm{As}$) which has to be fixed as a function of the radius of the target sphere, but for simplicity here we use the equation (\ref{prestress_curvature}) in order to compute the radius of the object that can be obtained at $\alpha=0.12.$

\begin{figure}
\begin{center}
\rotatebox{270}{
\begin{tikzpicture}[scale=1]
  \coordinate (c1) at (0,0);
  \draw[fill=blue,draw=blue,line width = 1.5mm] (-18*1.5708mm, 0mm) -- (18*2.4mm,0mm);
  \draw[fill=blue,draw=blue,line width = 1.5mm] (0cm,-18*2*1.5708mm) -- (0cm,18*2*1.5708mm);
   \coordinate (c1) at (105.22mm,0);
   \draw[fill=blue,draw=blue,line width = 0.1mm]
   ($(c1) + (180-31.25:102.83mm)$) arc ( 180-31.25:180+31.25:102.83mm) --
   ($(c1) + (180+31.25:101.33mm)$) arc ( 180+31.25:180-31.25:101.33mm) -- cycle;
   \coordinate (c1) at (-105.22mm,0);
   \draw[fill=blue,draw=blue]
   ($(c1) + ( -31.25:102.83mm)$) arc (      -31.25:+31.25:102.83mm) --
   ($(c1) + ( +31.25:101.33mm)$) arc (      +31.25:-31.25:101.33mm) -- cycle;
   \coordinate (c1) at (55.74mm,0);
   \draw[fill=blue,draw=blue]
   ($(c1) + (180-61.56:50.20mm)$) arc ( 180-61.56:180+61.56:50.20mm) --
   ($(c1) + (180+61.56:48.70mm)$) arc ( 180+61.56:180-61.56:48.70mm) -- cycle;
   \coordinate (c1) at (-55.74mm,0);
   \draw[fill=blue,draw=blue]
   ($(c1) + ( -61.56:50.20mm)$) arc (      -61.56:+61.56:50.20mm) --
   ($(c1) + ( +61.56:48.70mm)$) arc (      +61.56:-61.56:48.70mm) -- cycle;
   \coordinate (c1) at (40.60mm,0);
   \draw[fill=blue,draw=blue]
   ($(c1) + (180-90:31.92mm)$) arc ( 180-90:180+90:31.92mm) --
   ($(c1) + (180+90:30.42mm)$) arc ( 180+90:180-90:30.42mm) -- cycle;
   \coordinate (c1) at (-40.60mm,0);
   \draw[fill=blue,draw=blue]
   ($(c1) + ( -90:31.92mm)$) arc (      -90:+90:31.92mm) --
   ($(c1) + ( +90:30.42mm)$) arc (      +90:-90:30.42mm) -- cycle;
   \coordinate (c1) at (34.01mm,0);
   \draw[fill=blue,draw=blue]
   ($(c1) + (180-115.7:22.2mm)$) arc ( 180-115.7:180+115.7:22.2mm) --
   ($(c1) + (180+115.7:20.7mm)$) arc ( 180+115.7:180-115.7:20.7mm) -- cycle;
   \coordinate (c1) at (-34.01mm,0);
   \draw[fill=blue,draw=blue]
   ($(c1) + ( -115.7:22.2mm)$) arc (      -115.7:+115.7:22.2mm) --
   ($(c1) + ( +115.7:20.7mm)$) arc (      +115.7:-115.7:20.7mm) -- cycle;
   \coordinate (c1) at (30.81mm,0);
   \draw[fill=blue,draw=blue]
   ($(c1) + (180-137.9:15.85mm)$) arc ( 180-137.9:180+137.9:15.85mm) --
   ($(c1) + (180+137.9:14.35mm)$) arc ( 180+137.9:180-137.9:14.35mm) -- cycle;
   \coordinate (c1) at (-30.81mm,0);
   \draw[fill=blue,draw=blue]
   ($(c1) + ( -137.9:15.85mm)$) arc (      -137.9:+137.9:15.85mm) --
   ($(c1) + ( +137.9:14.35mm)$) arc (      +137.9:-137.9:14.35mm) -- cycle;
   \coordinate (c1) at (29.24mm,0);
   \draw[fill=blue,draw=blue]
   ($(c1) + (180-155.9:11.14mm)$) arc ( 180-155.9:180+155.9:11.14mm) --
   ($(c1) + (180+155.9: 9.64mm)$) arc ( 180+155.9:180-155.9: 9.64mm) -- cycle;
   \coordinate (c1) at (-29.24mm,0);
   \draw[fill=blue,draw=blue]
   ($(c1) + ( -155.9:11.14mm)$) arc (      -155.9:+155.9:11.14mm) --
   ($(c1) + ( +155.9: 9.64mm)$) arc (      +155.9:-155.9: 9.64mm) -- cycle;
   \coordinate (c1) at (28.54mm,0);
   \draw[fill=blue,draw=blue]
   ($(c1) + (180-169:7.3mm)$) arc ( 180-169:180+169:7.3mm) --
   ($(c1) + (180+169: 5.8mm)$) arc ( 180+169:180-169: 5.8mm) -- cycle;
   \coordinate (c1) at (-28.54mm,0);
   \draw[fill=blue,draw=blue]
   ($(c1) + ( -169:7.3mm)$) arc (      -169:+169:7.3mm) --
   ($(c1) + ( +169: 5.8mm)$) arc (      +169:-169: 5.8mm) -- cycle;
   \coordinate (c1) at (28.3mm,0);
   \draw[fill=blue,draw=blue]
   ($(c1) + (180-150: 3.9mm)$) arc ( 180-150:180+150: 3.9mm) --
   ($(c1) + (180+150: 2.4mm)$) arc ( 180+150:180-150: 2.4mm) -- cycle;
   \coordinate (c1) at (-28.3mm,0);
   \draw[fill=blue,draw=blue]
   ($(c1) + ( -177:3.9mm)$) arc (      -177:+177:3.9mm) --
   ($(c1) + ( +177: 2.4mm)$) arc (      +177:-177: 2.4mm) -- cycle;
   \filldraw[fill=blue,draw=blue] (18*2.40mm,0) circle (12pt);
\end{tikzpicture}
}
\caption{The planar grid designed to cover the sphere. The horizontal straight line will fully cover the ecuator while the lower and upper parts will cover the North and South hemispheres, respectively. Notice the slight modification of the length for small arcs near the South pole needed in order to keep the relaxed structure attached through the filled round dot (with a characteristic size larger than the lateral dimensions of the curved ribbons) to the substrate during the under-etching process.}
\label{design_picture}
\end{center}
\end{figure}
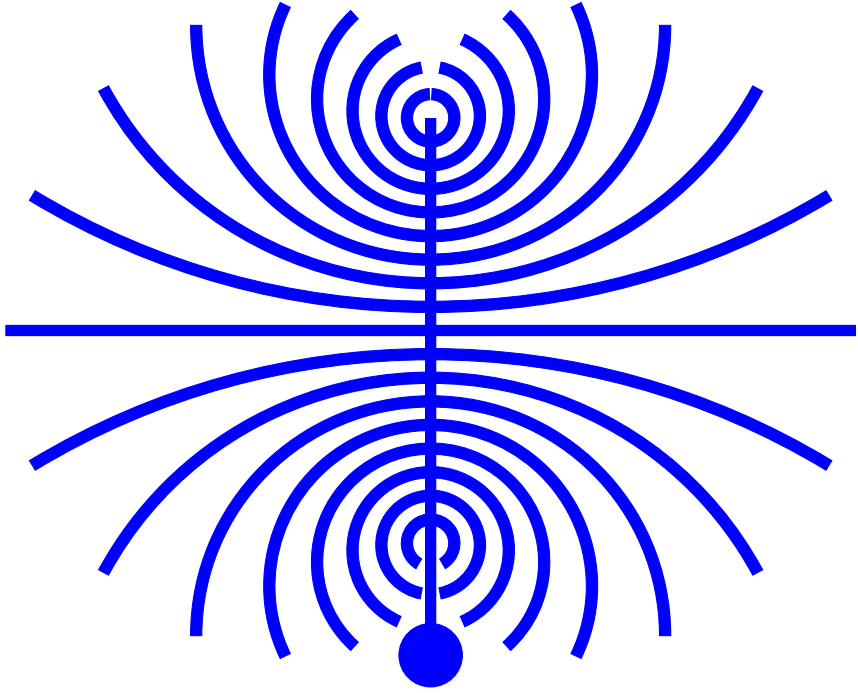

\subsection{Design and fabrication}

\begin{figure}[ht!]
\centering
\includegraphics[width=0.9 \textwidth]{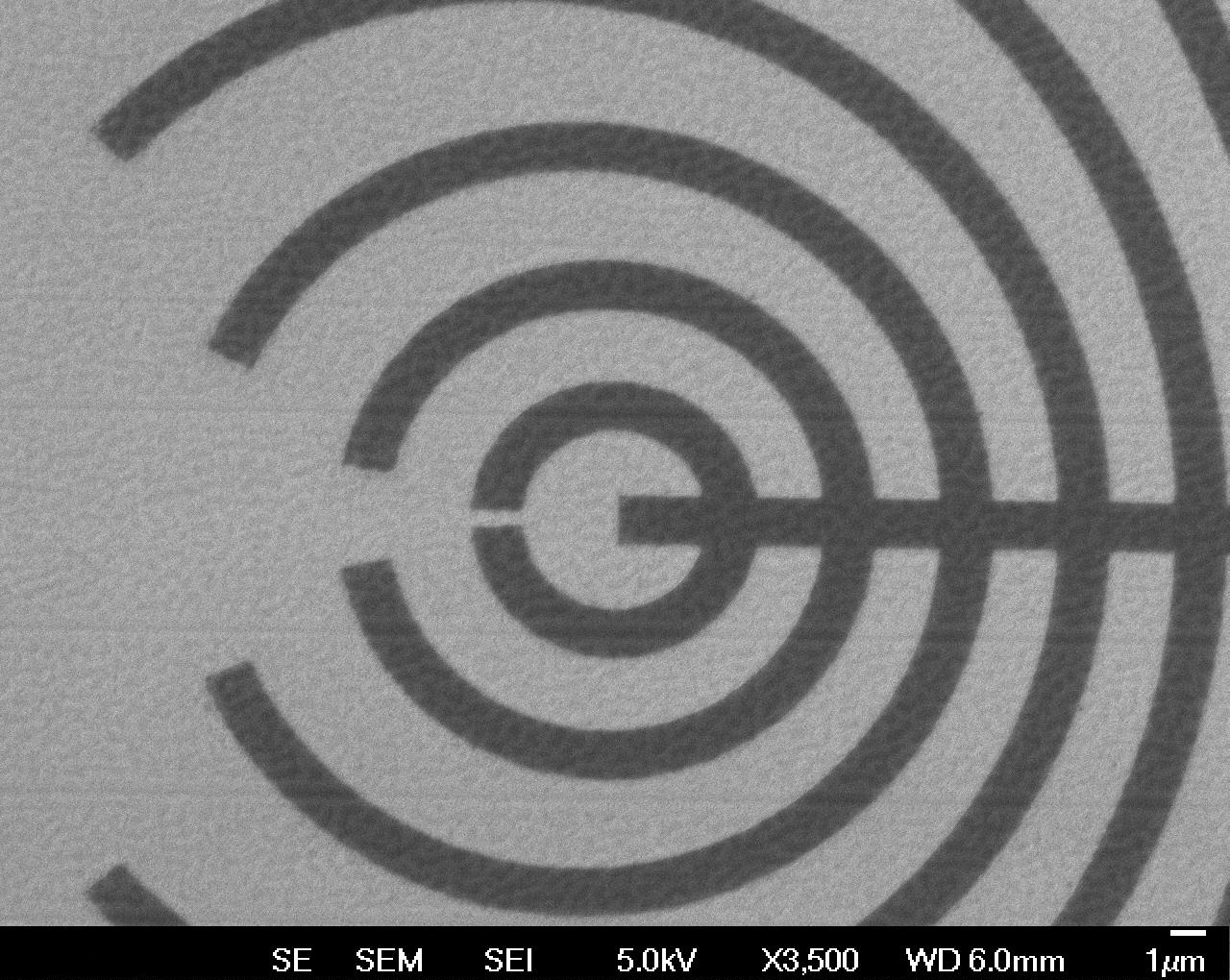}
\caption{SEM image of the sample after the developement process before the reactive ion etching (RIE).}
\label{litho}
\end{figure}

\begin{figure}[ht!]
\begin{center}
\includegraphics[width=0.8 \textwidth]{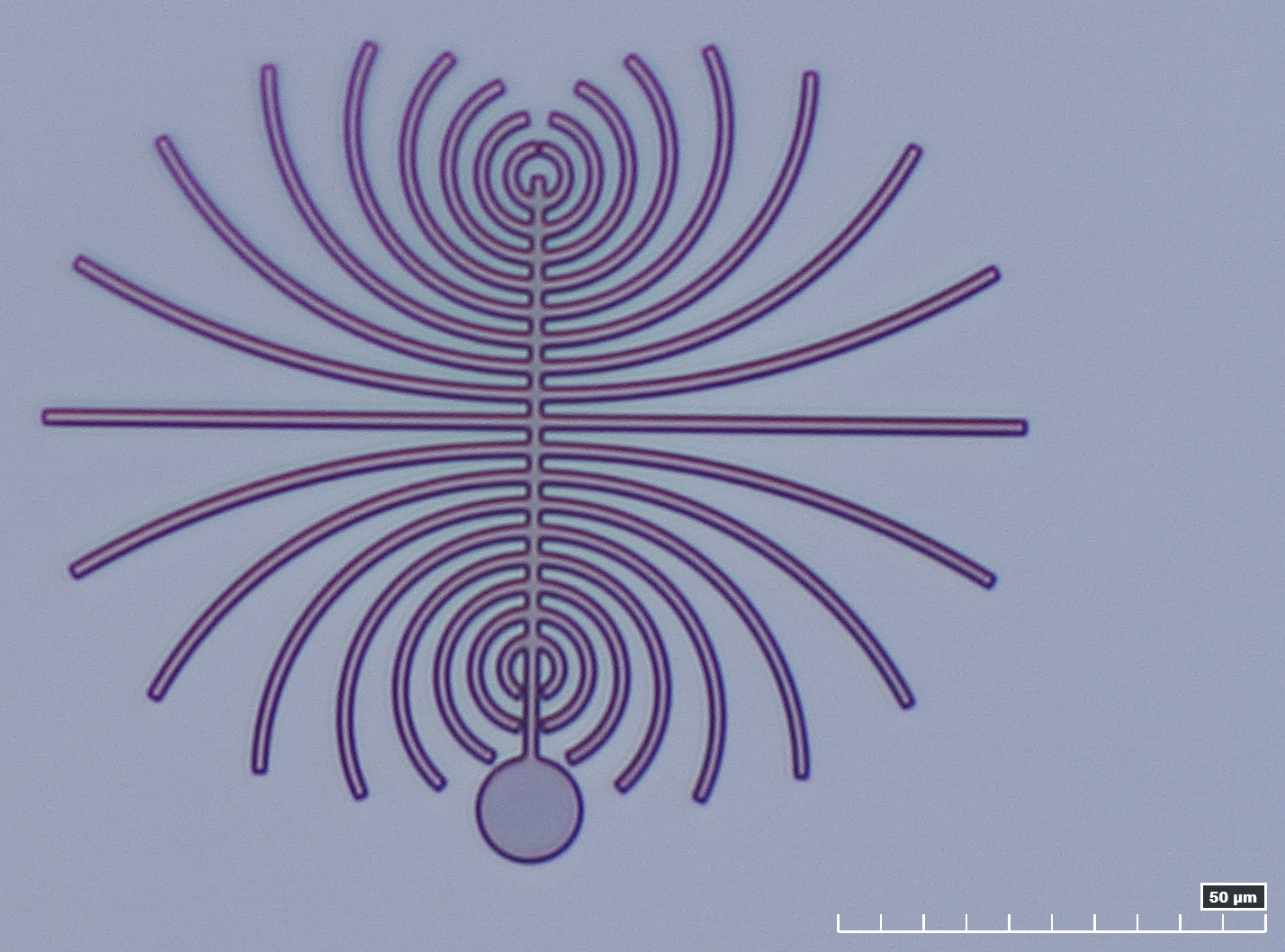}
\end{center}
\includegraphics[width=0.5 \textwidth]{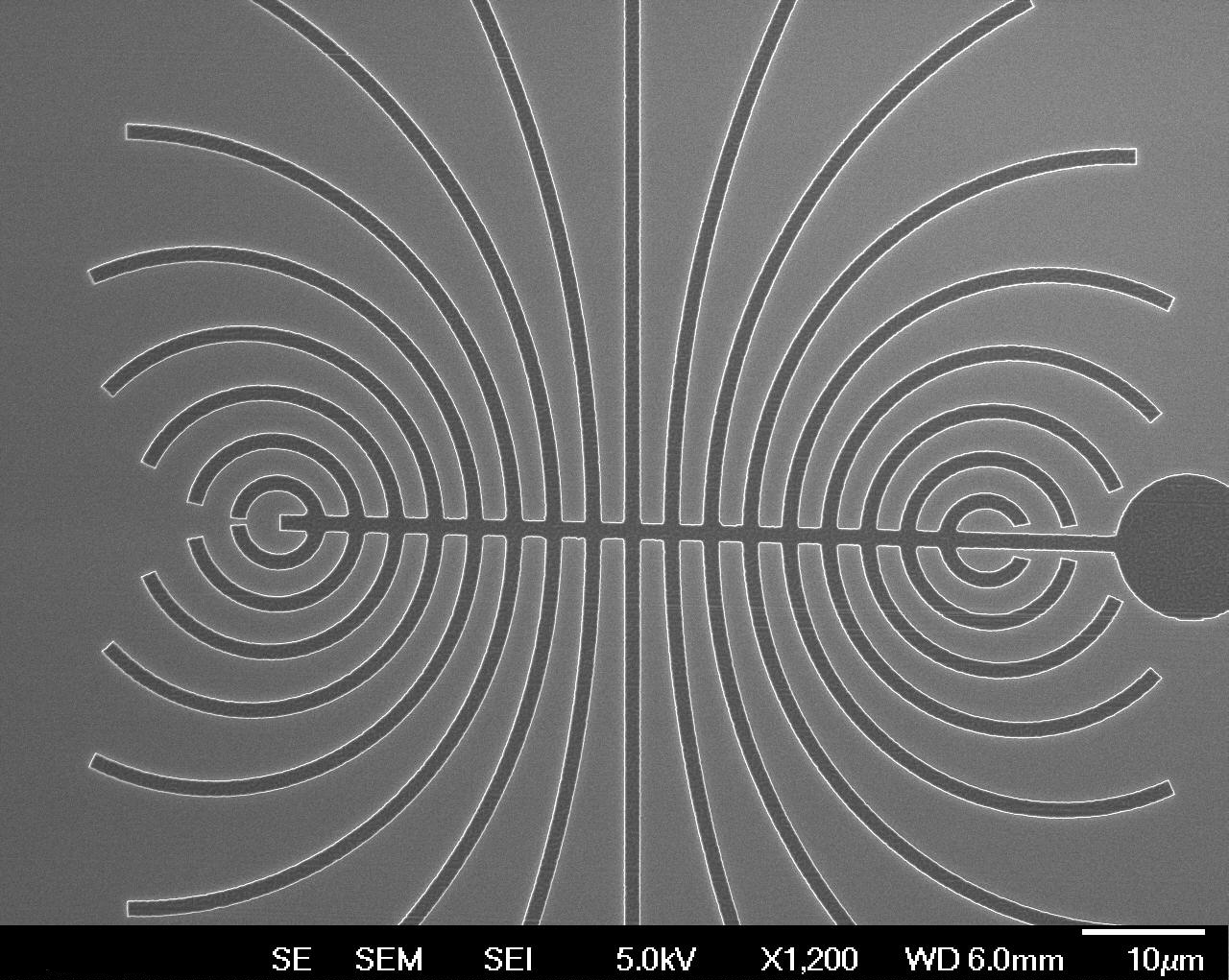}
\includegraphics[width=0.5 \textwidth]{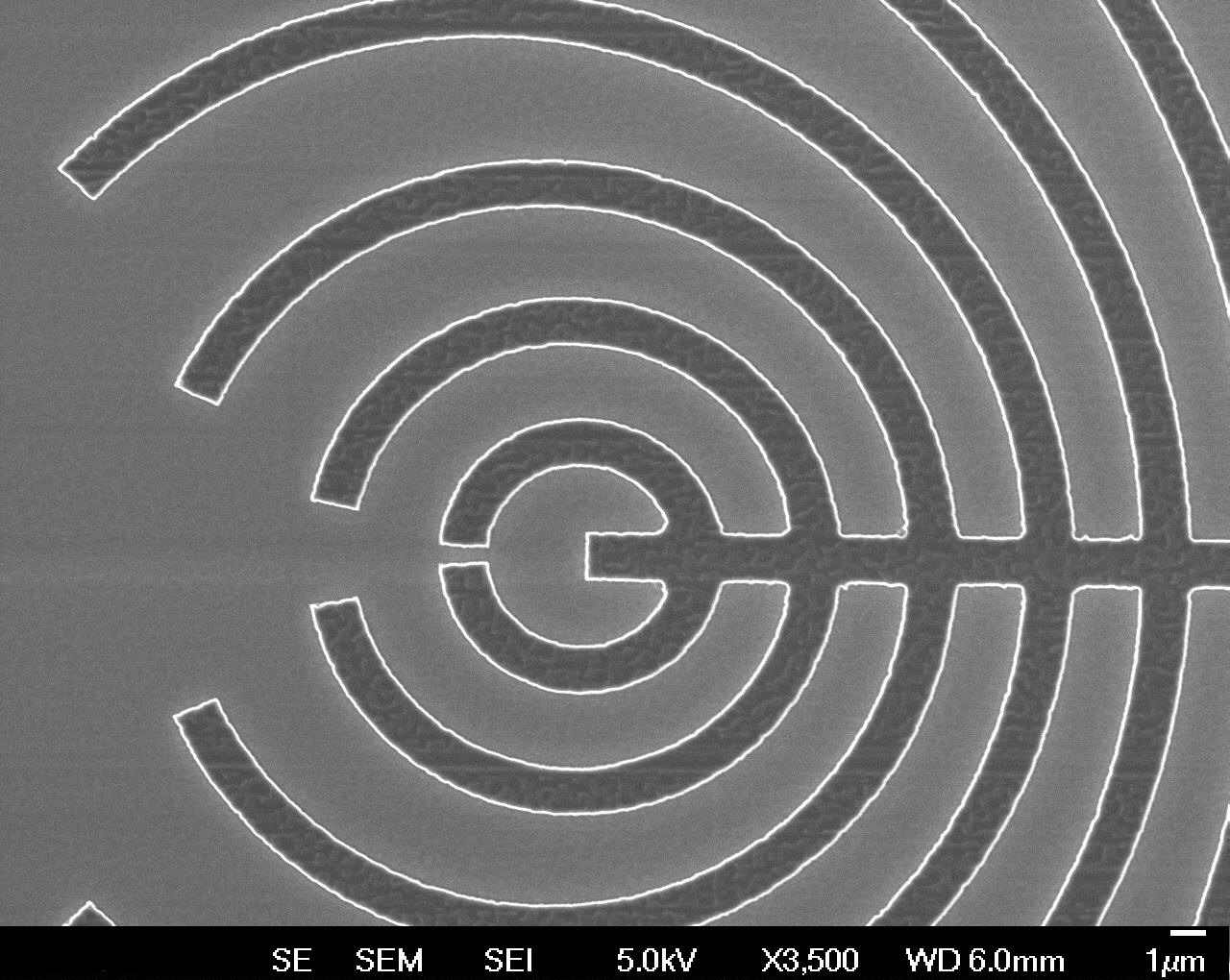}
\caption{Optical microscope view of the structure obtained after etching the multilayer material, still maintained attached to the substrate by the sacrificial layer. }
\label{optical}
\end{figure}

In order to cover the sphere of radius $R_*$ with constant latitude ribbons we notice that the radius of the ribon at latitude $\theta$ is $R_*\cos\theta$ and their geodesic curvature, which is exactly the inverse of tha planar design radius, is constant and equal to $\kappa(\theta)=\frac{1}{R_*}{\tan\theta}.$ For simplicity, we chose the width of all ribbons equal to $1.5$
$\mu$m (for visual comfort, the actual scale in Figure \ref{design_picture} is not the same as the implemented design in \ref{optical}) and design a geodesic half-circle to ensure connectivity between the constant latitude ribbons. Using the intersection of the two straigt lines in figure \ref{design_picture} as the origin of the coordinate system in the plane, positions of the 8 pairs of symetric arcs corresponding to constant lattitude arcs located at $\pm\frac{n\pi}{18}$ ($n=1,\ldots,8$) in the North and South hemi-sphere. Their corresponding centers radii and angular extensions are 
\begin{equation}
C_n^\pm = (0,\pm R\left(\frac{n\pi}{18}+\frac{1}{\tan\theta}\right),\quad
R_n = R/\tan\theta,\quad
\theta_n = \pi\sin(\frac{n\pi}{18}).
\end{equation}

\begin{figure}[ht!]
\centering
\includegraphics[width=0.8 \textwidth]{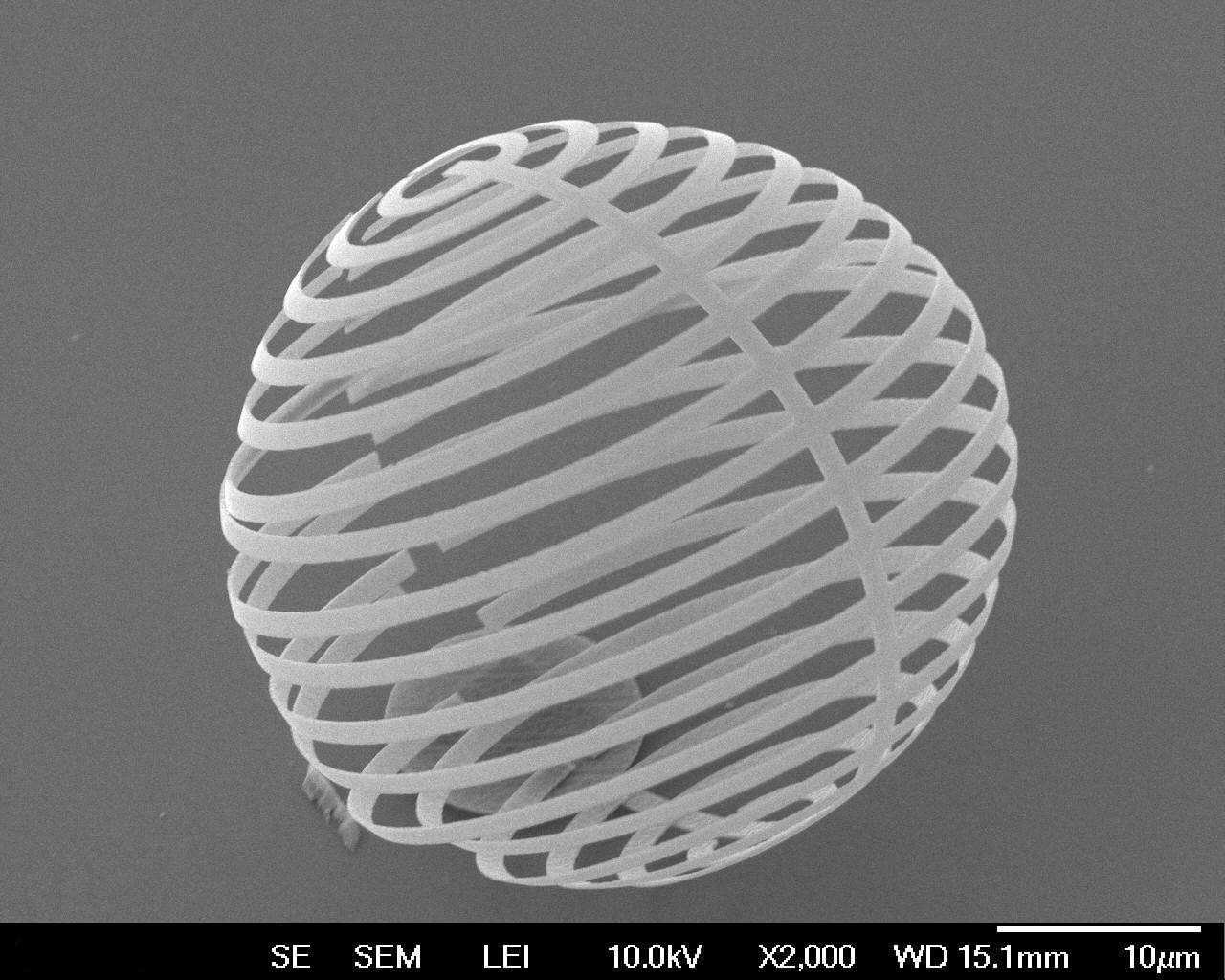}
\caption{The relaxed shape confirms that the geodesic curvature design relax as expected into constant latitude circles on the sphere.}
\label{confirmation}
\end{figure}

Fabrication of the design illustrated in Figure \ref{design_picture} involve several steps : we start by the deposition of a 90 nm thick $\textrm{SiO}_2$ hard mask followed by the deposition of a 130 nm thick negative resist film (AR-N7520.07). Next step is electron beam lithography performed by using a modified SEM (FEI Inspect F) and the RAITH Elphy Quantum software. The result after the developement of the lithographic process is illustrated in Figure \ref{litho}. The reactive ion etching (RIE) is then performed in order to transfer the pattern into the silica mask and then into the multilayer structure. The result of this process is represented in Figures \ref{optical} (both optical microscope and SEM images). At this step, the structure is still attached on the sacrificial layer but the lateral relaxation of the pre-stress in the bilayer material takes place. Despite the small width ($1.5$ $\mu$m) and the ultra-small thickness ($205$
nm) the axial pre-stress is still present in the structure and will be released only during the process of under-etching of the sacrificial layer. In order to keep the relaxed structure attached to the substrate the radius of the attachement circle (designed in the lower part of Figure \ref{design_picture}) have to be slightly larger then the width of various arcs of the design.

Next, the under-etching is performed using diluted $\textrm{FeCl}_3$ to selectively remove the InGaAs sacrificial layer so as to release the pre-stress in the multilayer. Successive $\textrm{H}_2\textrm{O},$ acetone and methanol rising baths were performed before a $\textrm{CO}_2$ supercritical drying step, needed in order to circumvent the mechanical actions induced by the surface tension at liquid/solid interfaces. As expected, the fully relaxed structure covers the surface of the sphere with constant latitutde ribbons, with only small alignement defects at the ends. Obviously, a large variety of different designs can be implemented but, as already stated, techological limitations associated with the photolithographic process (sharp angles) do not allow all of these designs to be sucessfully implemented.

\section{Conclusions and perspectives}

\begin{figure}[ht!]
	\begin{center}
    \includegraphics[width=0.8 \textwidth]{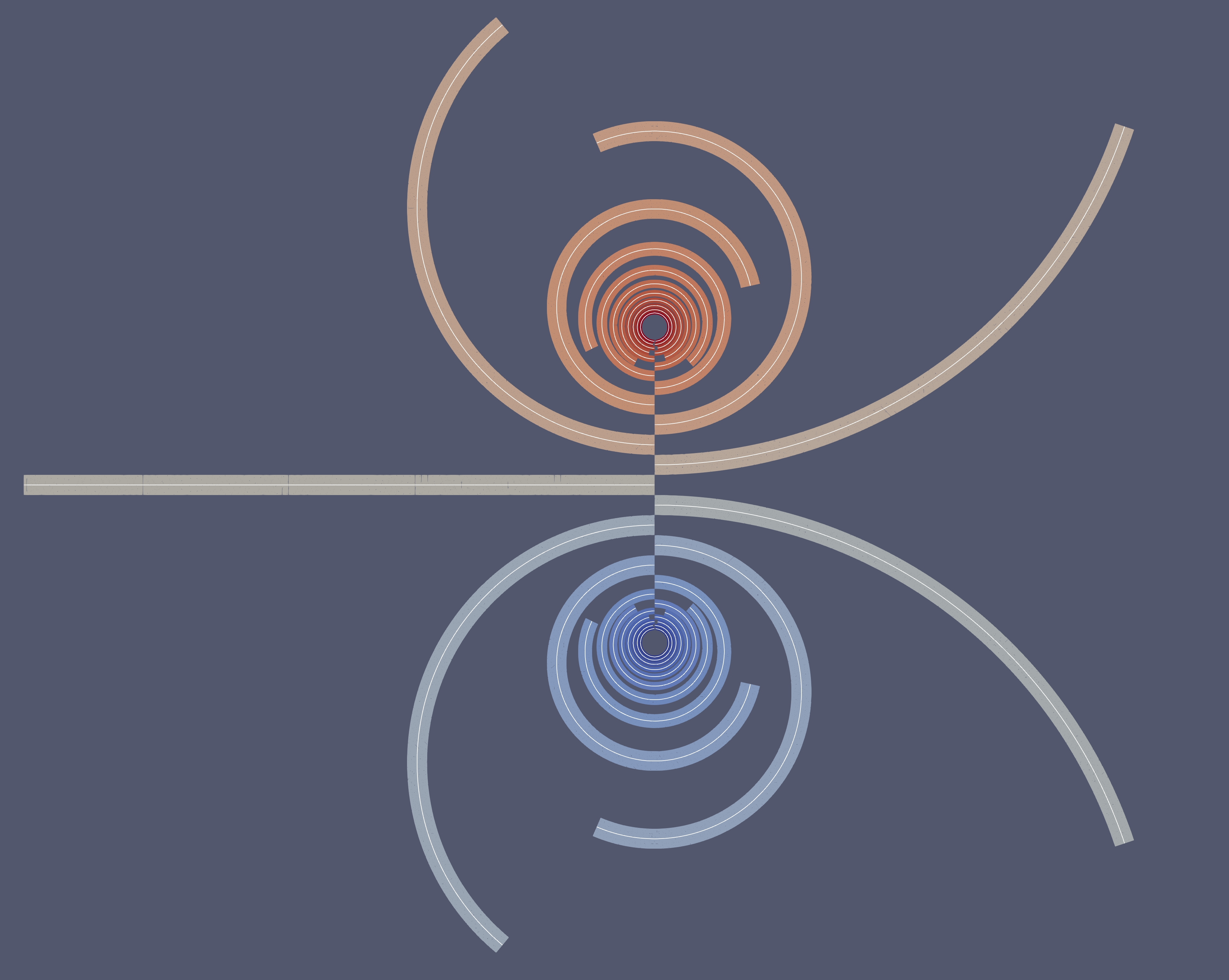}
	\end{center}
	\includegraphics[width=0.5 \textwidth]{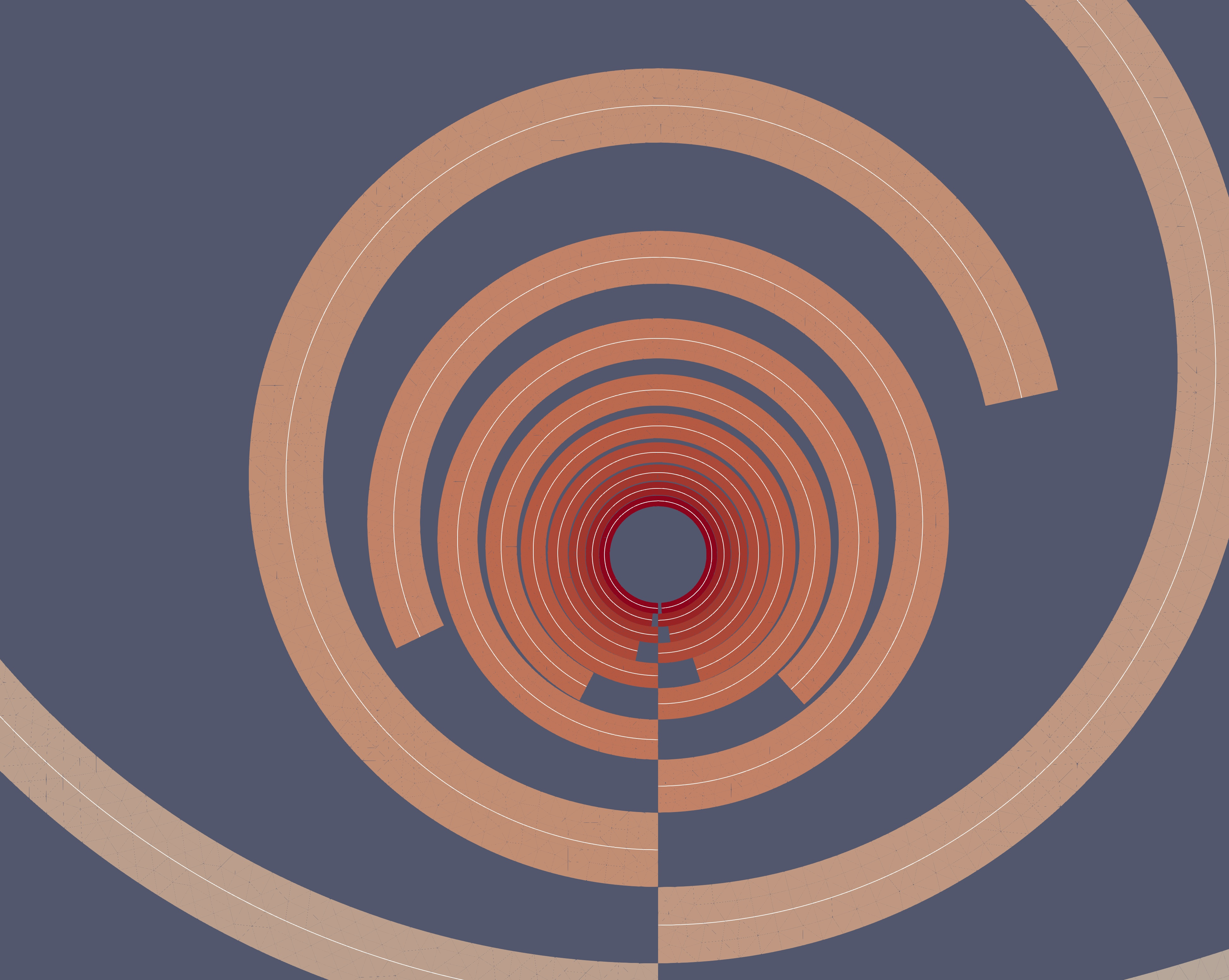}
	\includegraphics[width=0.5 \textwidth]{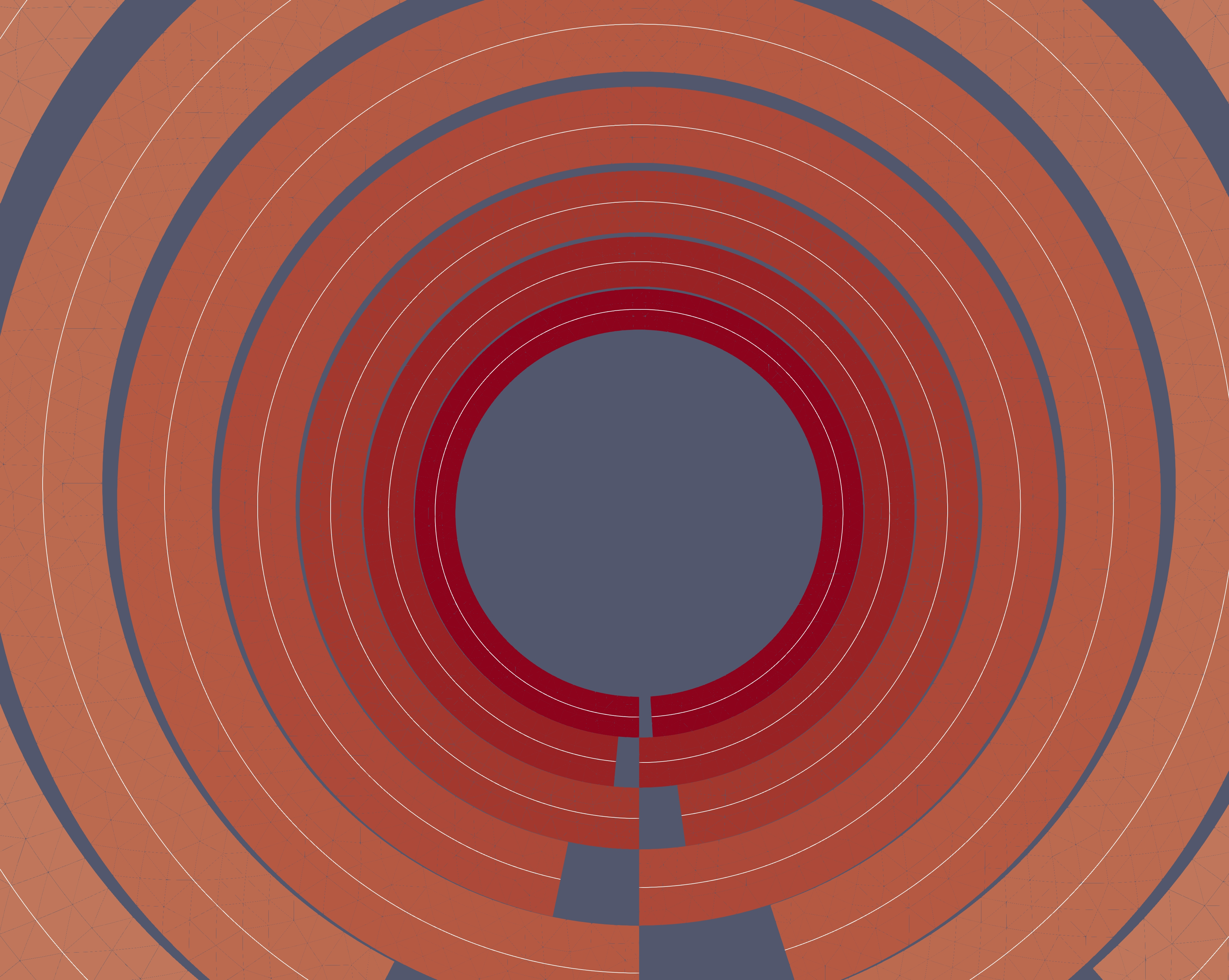}
	\caption{Top: Optimal covering of the sphere for $\delta=10^{-2}$. Alternative distribution of the constant latitude arcs on the left (respectively right) part of the initial design present in Figure \ref{design_full} to avoid   sharp angles along the vertical symmetry axis. Bottom: zoom on the central zone.  }
		
	\label{solution}
\end{figure}

The fabrication of nano-shells, is in itself a technological challenge as it encompass the traditional planar technology. One way to obtain such structures is to release the pre-stressed nano-plates, fabricated by layer-by-layer deposition, to obtain a target shell geometry. The presence of several geometrical and technological restrictions can be circumvent by the use of three-dimensional thin ribbons in order to cover a desired surface. The geodesic curvature plays an fundamental role for the design of both the geometry of ribbons that cover arbitrary surfaces starting from planar structures and the pre-stress needed to obtained them. The main result shows that, for multilayered structures with weak-transversal homogeneity, if the curvature of the planar ribbon is equal to the geodesic curvature of the supporting curve then there exists a pre-stress such that a small-width and small-thickness planar ribbon relaxes toward a 3D ribbon covering the surface along the suporting curve. We illustrate our theoretical results by the design and fabrication of a partial cover of the sphere with constant-latitude ribbons starting from a planar design containing arcs with constant curvature and a bilayer semi-conductor bilayer material with controled composition ($\textrm{In}_{0.88}\textrm{Ga}_{0.12}\textrm{P}/\textrm{InP}$). 

Extensions of these results to obtain a complete cover of the sphere are limited by the resolution of the lithographic process, difficult to implement at very sharp angles. A solution to overcome this technological drawback of the (sharp angles) lithographic process is illustrated in Figure \ref{solution}. Here, in order to avoid the sharp angles between successive ribbons located at constant latitude we chose to design alternative left and right constant radius arcs corresponding to successive constant latitude ribbons. The ideal picture in Figure \ref{solution} does not include a small vertical segment, which is nedded in order to attach the constant curvature arcs to the structure. In fact, the design in Figure \ref{solution} contains exactly the same arcs as that of the Figure \ref{design_full} but their positions are such that sharp angles along the vertical symmetry line in Figure \ref{design_full} are avoided. However, a closer look to the design in Figure \ref{solution}, reveals very small distances between successive ribbons (also present in the initial design in Figure \ref{design_full} between large latitude ribbons.

The general results in \cite{danescu2021} provide solutions to both partial and total covers of other non-developable (orientable or not) three-dimensional surfaces as the torus and the Mobius ribbon, extending the classical setting of isometric transformations. We mention here two interesting extensions: the first one concerns the class of arbitrary transversal homogeneities (and not only weak transversal homogeneities) in which case one has to adapt the general setting in \cite{danescu2020}. The second perspective concerns the fabrication of more complex geometries which require not only bending but also torsion, a problem already discussed in \cite{danescu2021} which is dependent to more complex (not only hydrostatic) pre-stress. Controled spatial modulation of the pre-stress, and in particular including controlled shear still remains a technological landmark at the nano-scale.       

\bibliographystyle{spbasic}
\bibliography{design}

\end{document}